\documentclass[journal=jacsat,manuscript=article]{achemso}

\usepackage[version=3]{mhchem} % Formula subscripts using \ce{}
\usepackage{placeins}
\usepackage{amsmath}
\usepackage{booktabs}       % professional-quality tables
\usepackage{amsfonts}       % blackboard math symbols
\usepackage{nicefrac}       % compact symbols for 1/2, etc.
\usepackage{microtype}      % microtypography
\usepackage{xcolor}         % colors
\usepackage{amsmath}
\usepackage{graphicx}
\usepackage{mathtools, amssymb, lipsum, nccmath} 
\usepackage{placeins}
\usepackage{enumitem} 
\usepackage{siunitx}
\usepackage{nameref}
\usepackage{lineno}
\usepackage{hyperref}
\newif\ifshowchanges
\author{Jared Solovei}
\author{Sakshi Agarwal}
\author{Aubrey Toland}
\affiliation[Georgia Tech]
{School of Materials Science and Engineering, College of Engineering, Georgia Institute of Technology, Atlanta, Georgia 30318, United States}
\author{Jeffrey G. Ethier}
\author{Vikas Varshney}
\affiliation[AFRL]
{Foundational Technologies Directorate, Air Force Research Laboratory, Wright-Patterson Air Force Base, Ohio 45433, United States}
\author{Lisa M. Hall}
\affiliation[OSU ChemE]
{William G. Lowrie Department of Chemical and Biomolecular Engineering, The Ohio State University, Columbus, Ohio 43210, United States}
\author{Wolfgang Windl}
\affiliation[OSU MSE]
{Department of Materials Science and Engineering, The Ohio State University, Columbus, Ohio 43210, United States}
\author{Rampi Ramprasad}
\email{rampi.ramprasad@mse.gatech.edu}
\affiliation[Georgia Tech]
{School of Materials Science and Engineering, College of Engineering, Georgia Institute of Technology, Atlanta, Georgia 30318, United States}
\title[LLM Prediction of Polymer Atomic Oxygen Erosion]
  {Leveraging Large Language Models to Predict Atomic Oxygen Erosion of Polymers in Low Earth Orbit}

\begin{document}



%%%%%%%%%%%%%%%%%%%%%%%%%%%%%%%%%%%%%%%%%%%%%%%%%%%%%%%%%%%%%%%%%%%%%
\begin{abstract}
Low Earth orbit (LEO), home to most active satellites, is projected to undergo a debris-fragmentation cascade known as Kessler Syndrome. In light of this eventuality, it is imperative to maximize the operational lifetime of LEO satellites. Instrumental to this goal is the development of next-generation polymeric materials with improved resistance to atomic oxygen (AO) erosion. Because genuine LEO exposure experiments are prohibitively expensive and LEO conditions are difficult to reproduce in the laboratory, data-driven approaches offer a promising alternative for accelerating materials discovery. Here, we develop a fine-tuned large language model (LLM) based on OpenAI’s GPT-4o to predict the AO erosion yield ($E_y$) of polymers using a curated dataset of 201 measurements compiled from multiple NASA LEO exposure missions. The model achieves an order of magnitude error (OME) of 0.154 and a log-scaled coefficient of determination (log-R\textsuperscript{2}) of 0.719, outperforming conventional machine learning approaches while eliminating the need for extensive feature engineering. Notably, incorporating macroscopic descriptors such as film thickness and layering did not improve predictive performance, suggesting that AO erosion is governed primarily by chemical and environmental variables rather than bulk structural characteristics. This finding underscores the importance of chemically informed polymer design in developing materials with improved AO resistance, while demonstrating the ability of modern LLMs to effectively extract chemically meaningful relationships even when data is limited. Beyond its predictive accuracy, the proposed framework provides a scalable platform for rapidly screening candidate polymers, reducing reliance on costly spaceflight experiments. As additional LEO datasets become available, integrating high-fidelity flight data with complementary ground-based experiments and simulation results through data-fusion strategies, such as multitask learning, offers a promising pathway toward more accurate predictive models and a deeper mechanistic understanding of AO erosion in space environments.
\end{abstract}

\section{Introduction}

The low Earth orbit (LEO) region extends from roughly 160 km to 2000 km in altitude and is home to nearly 80\% of active satellites.\cite{doi:10.34133/2022/9865174} This highly populated region is the epicenter of an emerging space debris problem that, if not properly addressed, will cause considerable harm to the space industry, weather forecasting and satellite-based communication. The present LEO environment is sufficiently unstable that collisions between existing objects will become the dominant source of debris generation, even in the absence of new launches. However, the arrival of this collision-driven, self-perpetuating environment may occur over decades or centuries, depending on the care with which future missions are conducted.\cite{KESSLER199163, doi:10.1126/science.1121337}

Material degradation in LEO is driven by extreme conditions, characterized by a high flux of diradical atomic oxygen (AO), bombardment by solar photon radiation such as ultra-violet (UV), extreme UV (EUV) and X-rays, energetic charged particles and thermal cycling.\cite{NASA-HDBK-6024} These conditions contribute to loss of thermal control, instrument damage, and structural wear, all of which can prematurely end missions and contribute to the growing population of uncontrolled derelicts available to collide with and further populate the LEO debris field. For this reason, the development of the next generation of space polymers resistant to LEO conditions is of considerable interest to the scientific community and society at large. The current standard for protective polymers on spacecraft exteriors includes Kapton polyimide (Kapton HN and variants containing carbon or inorganic fillers) and back-surface-aluminized fluorinated ethylene propylene (FEP/Al).\cite{NASA-HDBK-6024,Snow2019MISSE7} Despite decades of research, no material has broadly supplanted Kapton or FEP as the standard space polymer. This lack of progress is most directly attributed to the immense difficulty and cost of conducting experiments in LEO.

Methods for testing polymer durability in LEO generally fall into one of three categories: flight experiments, ground-based testing and computational predictions. These methods quantify a sample's durability using the atomic oxygen erosion yield ($E_y$), defined as the volume of material lost per incident oxygen atom.\cite{NASA-HDBK-6024} The most reliable measurements of polymer $E_y$ come from NASA’s Materials International Space Station Experiment (MISSE) missions, in which samples are fixed to the International Space Station (ISS) at various orientations.\cite{deGroh2019MISSE28,deGroh2025PCE} While MISSE missions are the latest and most sophisticated of NASA’s $E_y$ experiments, several similar studies preceded them, such as the Long Duration Exposure Facility (LDEF) and Evaluation of Oxygen Interaction with Materials Experiment (EOIM) missions.\cite{clark1984ldef,whitaker1993ldef,gregory1991erosion,camber1993property,whitaker1993ao171,brinza1994eoim3,scialdone1995evaluation,koontz1995eoim3mass,koontz1995eoim3mission,kamenetzky1995eoim3polymers,leger1993eoim3overview,Slemp1985STS8Composites,Visentine1988STS8Results} Unfortunately, in-space erosion measurements are both expensive and inefficient; it is therefore prudent to focus on more cost-effective, scalable material evaluation techniques, using flight data as the ground-truth benchmark by which their predictions are assessed.

Ground-based AO erosion measurements offer a more accessible and scalable alternative to flight experiments. However, this approach possesses its own limitations, as lab conditions fail to adequately replicate the LEO environment, differing in oxygen directionality, energy, and flux, as well as the presence of charged species and UV radiation. Further complicating matters, the consistent and reliable generation of AO is itself a nontrivial task; as a result, ground-based testing often involves oxygen ions. These inconsistencies inevitably result in $E_y$ measurements that differ from those observed under genuine LEO conditions. \cite{miller2008differences,rutledge1994comparison,rutledge1998fep,stambler2011peace,cool1999gamma,tagawa2004synergistic,tagawa2008polymer,shimamura2009synergistic,banks2013hyperthermal,banks2012ash}

A third strategy avoids the limitations of both flight and ground-based testing by using computational tools to predict durability in LEO. These methods range from bespoke $E_y$ predictors involving ad hoc feature selection and regression\cite{Banks2009ErosionYieldPrediction,Iskanderova1995PolymerErosion} to Monte Carlo modeling and molecular dynamics simulations of AO erosion for a limited selection of space polymers.\cite{Banks2014MISSE6Kapton,deGroh1994AtomicOxygenUndercutting,Liu2010AtomicOxygenUndercutting,Wei2024ReactiveMolecularDynamics,Wei2024DamageDegradationMechanicalProperties,Rahnamoun2014ReactiveMolecularDynamics} These techniques vary in their complexity and reliability; however, what they all share is a resource- and labor-intensive  process focused on a select few cases.

Machine learning (ML) has emerged as a powerful, scalable material design tool, achieving notable success over the past decade.\cite{tran2024design,CHEN2021100595,Ramprasad2017} By leveraging existing data, ML enables the development of predictive models that can guide and accelerate materials discovery. However, conventional ML approaches often rely on a complex feature engineering process to represent materials with numerical fingerprints. Furthermore, these models tend to perform poorly in the low-data regime, limiting their utility in space-material design. To overcome such challenges, recent studies have explored the use of fine-tuned large language models (LLMs) specialized for specific chemical tasks.\cite{doi:10.1021/acsmaterialslett.5c00054,mishra2025foundationallargelanguagemodels,doi:10.1021/acscatal.3c04956} These models offer a more user-friendly, natural-language framework, leverage the base model's already embedded knowledge and have demonstrated strong performance when data is sparse.\cite{Zhang_2025,Jablonka2024}
\begin{figure}[H]
    \centering
    \includegraphics[width=1\linewidth]{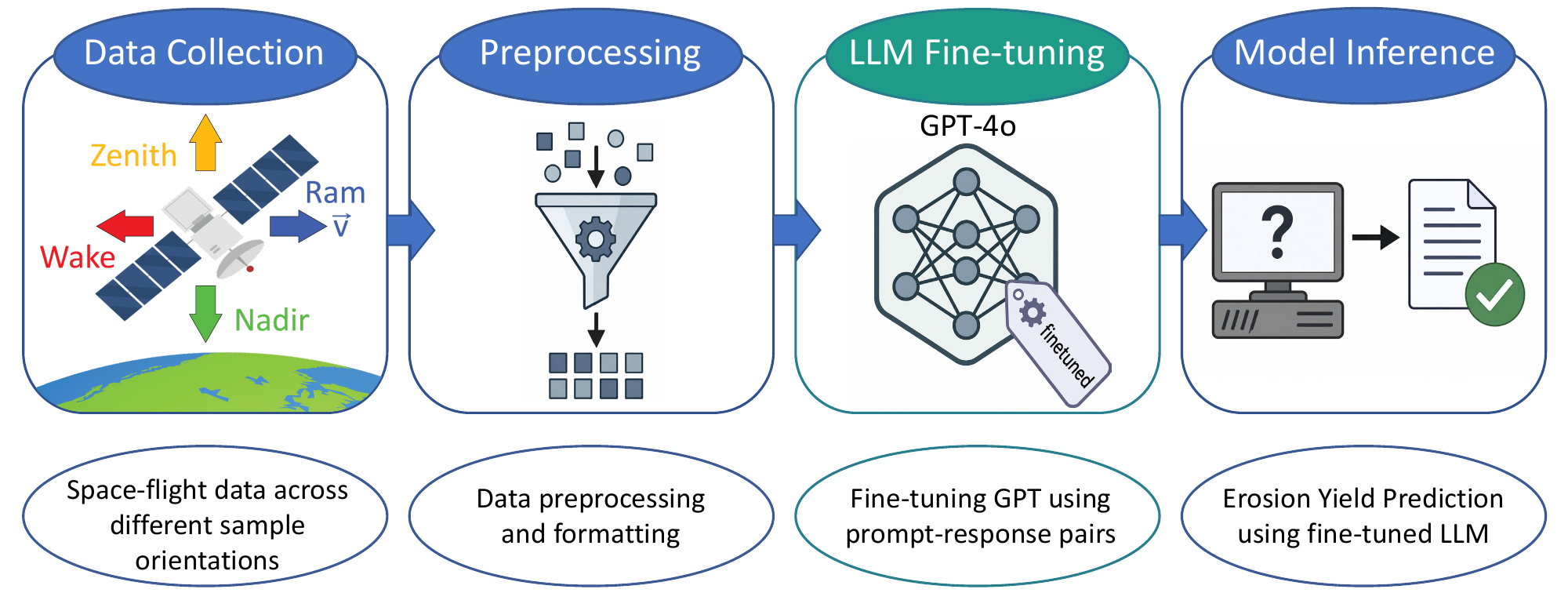}
    \caption{Workflow for polymer erosion yield ($E_y$) prediction. Space mission data are first collected across four spacecraft orientations: Ram, Nadir, Zenith, and Wake. The data is then preprocessed and formatted for training, followed by the fine-tuning of a GPT base model using prompt–response pairs. The resulting fine-tuned model is subsequently used to predict the $E_y$ for new polymer systems.}
    \label{fig1}
\end{figure}

Here, we present an approach for predicting the $E_y$ of polymers by fine-tuning OpenAI's GPT-4o model.\cite{GPT4oSystemCard2025,OpenAIModelOptimization2025} Our workflow begins with the collection of experimental $E_y$ data from a variety of NASA LEO missions. This is followed by a systematic preprocessing and curation procedure that yields a prompt-response dataset. The processed dataset is then used to fine-tune the GPT model, allowing it to learn statistical trends from example $E_y$ measurements and generalize to novel $E_y$ prediction tasks. This pipeline, illustrated in Figure~\ref{fig1}, results in a model that can serve as a versatile $E_y$ prediction tool applicable to a wide array of linear polymeric systems.

In order to verify the practical value of our approach, we also optimized and evaluated a traditional Gaussian process regression (GPR). The LLM outperformed the GPR benchmark in predicting AO erosion at high AO fluences and in generalizing to unseen chemistries, two capabilities central to identifying promising space polymer candidates. Through this approach, we establish a robust framework for rapidly screening polymers based on their AO erosion resistance, providing a practical means of space materials discovery capable of readily incorporating new data as it becomes available.

\section{Results and Discussion}
\subsection{Data Collection and Analysis}

\begin{figure}[H]
    \centering
    \includegraphics[width=1\linewidth]{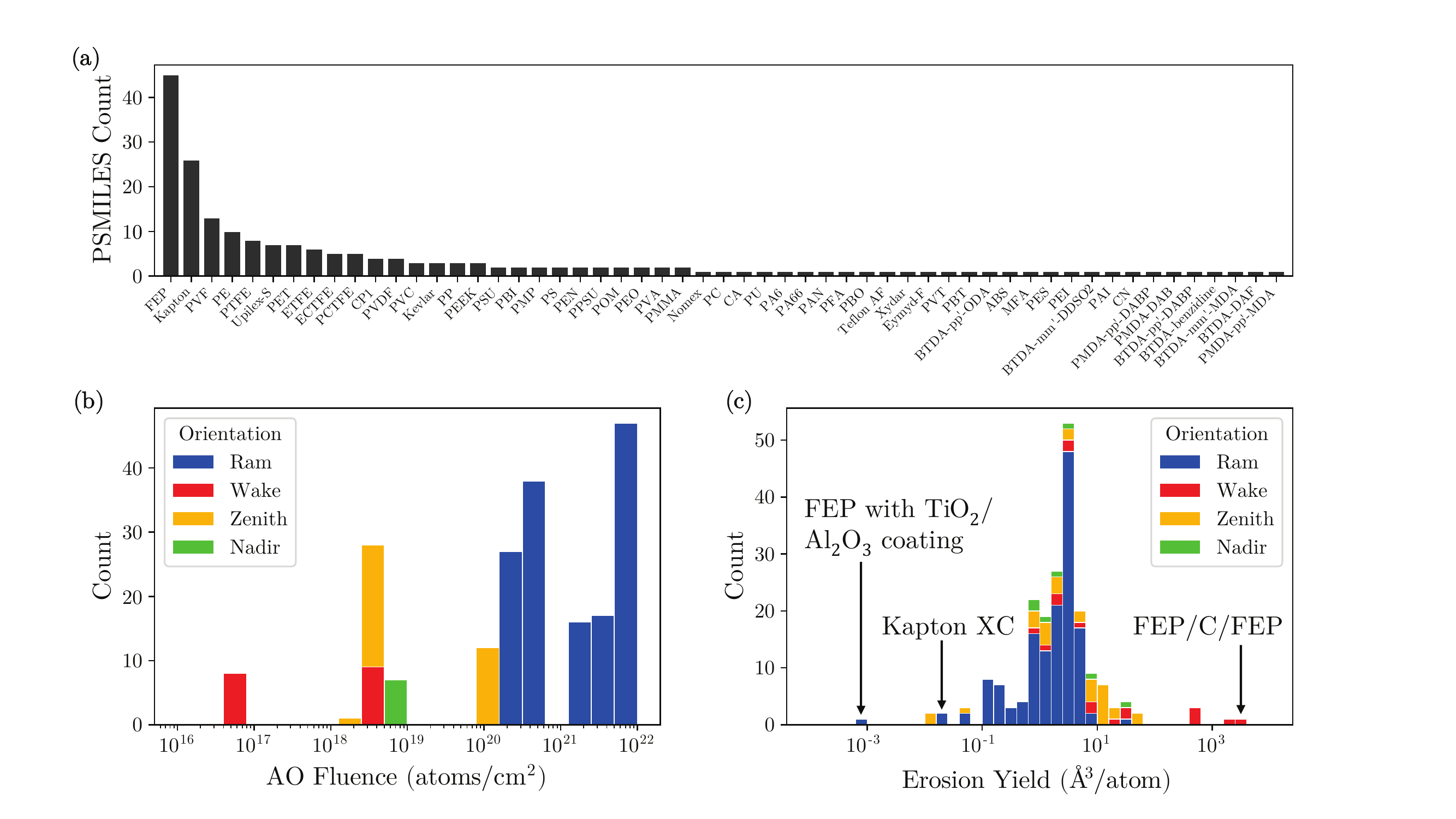}
    \caption{Erosion yield dataset overview. (a) Frequency of each polymer's assigned chemistry/PSMILES string in the dataset. For readability, each PSMILES string is referred to by its corresponding ``polymer name''. (b) Distribution of fluence values. (c) Distribution of erosion yield ($E_y$). The blue, red, yellow, and green colors depict the Ram, Wake, Zenith, and Nadir sample orientations, respectively.}
    \label{fig2}
\end{figure}

We collected 201 experimental $E_y$ measurements of polymeric materials from the MISSE, LDEF, EOIM III and STS-8 NASA missions conducted over several decades and multiple sample orientations,\cite{LEO_polymer_erosion_yield_dataset_2026} as outlined in Figure~\ref{fig1} and summarized in Figure~\ref{fig2} and Table~\ref{tab:dataset_summary}. These measurements correspond to 55 unique polymer chemistries, though these may incorporate various additives or fillers, possess different coatings or undergo different sample preparations. We encoded each polymer chemical type as a canonical PSMILES (Polymer Simplified Molecular Input Line Entry System) string based on its repeat unit(s) and using the canonicalization process described in prior work that generates a unique PSMILES string for a given repeat unit.\cite{kuenneth2023polybert} For 43 of these 55 chemistries, NASA provides drawings of the representative repeat units, while the remaining structures were found in the general literature or manufacturer datasheets based on the polymer’s chemical or trade name (additional detail is provided in the dataset’s “PSMILES source” and “PSMILES comment” columns).\cite{Banks2008ErosionYield,Visentine1988STS8Results} We note that there is a small subset of the original NASA samples for which valid PSMILES representations could not be assigned owing to structural complexity or incomplete chemical information. These materials were not included in the dataset to ensure a consistent and chemically meaningful representation for all polymers included in the study. For polymers with different repeat unit isomers—for example, polyvinyltoluene, in which the methyl group could be placed in the meta, ortho, or para position—only one isomer was used in the PSMILES structure. For copolymers, we created a canonical PSMILES string representing one repeat unit of each type bonded together. For example, a random AB copolymer of any composition is represented as an alternating AB copolymer (50\% A composition). In the case of Teflon AF 1601, which is made from a known 3:2 ratio of fairly small monomers, this ratio in an alternating structure was used to create the correct overall composition. For polyamide-imide Torlon 4203, its synthesis involves two types of diamines, leading to a long structural representation; we considered only the majority type of diamine resulting in a less complex PSMILES string. These choices were made to keep the PSMILES representations simple and compact while still encoding the chemical groups present. While representations like BigSMILES\cite{10.1021/acscentsci.9b00476} capture polymer architecture more explicitly and HAPPY\cite{Ahn2024} offers a compact, connectivity-focused representation of chemical groups, our current dataset contains a modest diversity of polymer chemistries and lacks instances of the same copolymer across varying compositions. Consequently, we adopted the standard canonical PSMILES approach, as it remains the most prevalent notation in the polymer informatics literature.

A related but distinct feature of our dataset is the 135 unique “polymer descriptions”, which provide the wording used to describe or name the material type in the reference. These descriptions provide or imply additional details beyond the repeat unit chemistry. The case of “Kapton HN” and “Kapton XC” provides an example of how different polymer descriptions can correspond to the same PSMILES string while also enabling a finer discrimination based on factors such as fillers, additives and commercial grades. Since the polymer chemistry in both samples is “Kapton”, these are assigned the canonical PSMILES string encoding poly(4,4’-oxydiphenylene pyromellitimide). However, Kapton XC incorporates conductive carbon particles not found in Kapton HN,\cite{dupont2006kaptonxc} a difference recognized within domain literature but omitted in the reference. Often, further details about these differences, such as filler volume fraction, were not available; thus, this information is included only in the “polymer description” of the dataset. Similarly, information pertaining to the coating applied to a sample is captured by one of the 16 text-based “coating descriptions”. These two features are clear instances of valuable natural-language descriptors that resist reliable, machine-readable representation.

Figure~\ref{fig2}(a) shows the frequency with which each PSMILES string appears in the dataset; two polymers, namely Kapton and FEP, make up more than a third of the entries. Despite this apparent chemical imbalance, these repeated measurements vary in environmental conditions and macroscopic film and material characteristics, providing insight into potential chemistry-independent influences on erosion. These descriptors include coating description, sample orientation, number of thin-film layers, the thickness of each layer, mission duration, AO fluence, and solar exposure (Table~\ref{tab:dataset_summary}). In particular, thin-film specimens vary in both layer count and per-layer thickness, with many incorporating protective coatings on the front, back, or interlayer surfaces. Likewise, samples flown on different NASA missions or exposed in different orientations are subject to distinct orbital environments, explaining variations in cumulative atomic oxygen (AO) bombardment and solar radiation exposure. 

Among the environmental descriptors, AO fluence provides a measure of the primary driver of in-orbit degradation. Defined as the time-integrated atomic oxygen (AO) flux, it quantifies the cumulative AO bombardment experienced by a sample and is generally estimated using a Kapton witness sample. A secondary descriptor is solar exposure, reported in equivalent sun hours (ESH), which quantifies cumulative exposure to high-energy electromagnetic radiation, solar wind and energetic particles. Solar radiation is believed to act synergistically with AO, further exacerbating material degradation. Collectively, each sample in our dataset is characterized by its polymer description, representative PSMILES string, coating description, orientation relative to the spacecraft velocity vector, AO fluence, solar exposure (ESH), mission duration, number of thin-film layers, and the thickness of each layer. A summary of the complete chemical, configurational, and environmental feature set is provided in Table~\ref{tab:dataset_summary}.

\begin{table}[H]
\centering
\caption{Description and summary statistics of the polymer atomic-oxygen erosion-yield dataset.}
\label{tab:dataset_summary}

\small
\renewcommand{\arraystretch}{1.18}
\setlength{\tabcolsep}{5pt}

\begin{tabular}{
@{}
p{0.20\textwidth}
p{0.41\textwidth}
p{0.29\textwidth}
@{}
}
\toprule
\textbf{Descriptor}
& \textbf{Description}
& \textbf{Data summary} \\
\midrule

Dataset size
& Total number of measurements
& 201 measurements \\
\addlinespace[3pt]
\cmidrule(lr){1-3}
\addlinespace[3pt]

Polymer description
& Textual, literature-derived description of the polymer identity
&  135 unique strings \\
\addlinespace[3pt]
\cmidrule(lr){1-3}
\addlinespace[3pt]

PSMILES
& Canonical string encoding the polymer chemistry assigned to each sample
& 55 unique chemistries, each assigned a ``polymer name'' for ease of reference. \\
\addlinespace[3pt]
\cmidrule(lr){1-3}
\addlinespace[3pt]

Coating description
& Textual description of coating applied to the sample, if any
& 16 unique coating descriptions \\
\addlinespace[3pt]
\cmidrule(lr){1-3}
\addlinespace[3pt]

Orientation
& Direction relative to the velocity vector
  (Ram, Nadir, Wake, Zenith)
& 145 ram-facing; 56 non-ram-facing \\
\addlinespace[3pt]
\cmidrule(lr){1-3}
\addlinespace[3pt]

AO fluence
& Time-integrated atomic-oxygen flux,
  representing total AO bombardment
& $4.46 \times 10^{16} - 8.99 \times 10^{21}$
  AO\,cm$^{-2}$ \\
\addlinespace[3pt]
\cmidrule(lr){1-3}
\addlinespace[3pt]

Solar exposure
& Exposure to solar radiation, reported in
  equivalent sun hours
& $7.9 - 11$\,200 ESH \\
\addlinespace[3pt]
\cmidrule(lr){1-3}
\addlinespace[3pt]

Mission time
& Duration of spaceflight exposure
& $0.00477 - 5.8$ years \\
\addlinespace[3pt]
\cmidrule(lr){1-3}
\addlinespace[3pt]

Layers$^{a}$
& Number of thin-film layers in the sample
& $1 - 24$ layers \\
\addlinespace[3pt]
\cmidrule(lr){1-3}
\addlinespace[3pt]

Thickness$^{a}$
& Thickness of each film layer
& $0.0127 - 0.737$ mm \\
\addlinespace[3pt]
\cmidrule(lr){1-3}
\addlinespace[3pt]

Mission
& NASA mission associated with each sample;
  not used as a model feature
& 13 missions \\
\addlinespace[3pt]
\cmidrule(lr){1-3}
\addlinespace[3pt]

$E_y$
& Atomic-oxygen erosion yield; prediction target
& $0.000799 - 2790$
  \si{\angstrom\cubed\per atom} \\

\bottomrule
\end{tabular}

\vspace{4pt}

\begin{minipage}{0.90\textwidth}
\footnotesize
$^{a}$ Recorded in the dataset and assessed in the feature-ablation study,
but excluded from the final predictive models.
\end{minipage}

\end{table}

Figure~\ref{fig2}(b) establishes the clear dependence of AO fluence on sample orientation, with Ram-facing samples receiving the heaviest bombardment and Wake-facing samples receiving significantly less. This is physically intuitive, since Ram-oriented surfaces are approximately normal to the satellite’s velocity, maximizing directional collisions with AO. Wake-oriented samples, by contrast, face opposite the direction of motion and are therefore subjected to fluences orders of magnitude lower than those of their Ram-facing counterparts. By extension, Zenith- and Nadir-facing samples, whose surfaces lie parallel to the direction of motion, are exposed to AO fluences between the Ram- and Wake-oriented extremes. In this way, the four orientations provide us a coarse-grained means of discriminating between high (Ram), moderate (Nadir and Zenith), and low (Wake) levels of AO bombardment.

Figure~\ref{fig2}(c) serves to demonstrate the apparent clustering by orientation within the $E_y$ distribution. In particular, Ram- and Wake-oriented samples predominantly occupy the low and high ends of the distribution, respectively, whereas Zenith- and Nadir-oriented samples fall between the two extremes. Together with the orientation--fluence relationship made evident by Figure~\ref{fig2}(b), the grouping of orientations in Figure~\ref{fig2}(c) suggests an inverse relationship between AO fluence and $E_y$. The observation that samples under greater cumulative AO bombardment exhibit lower $E_y$ values has been documented in several prior studies and is attributed to the formation of a protective layer of AO-resistant ash that accumulates on the sample's surface as its oxidizable components are eroded away.\cite{banks2012ash} Importantly, even polymers without intentionally added fillers or protective coatings exhibit this effect to some extent. Representative of this effect is the Wake-oriented sample of FEP with an interlayer carbon coating (FEP/C/FEP), shown in Figure~\ref{fig2}(c), which has an $E_y$ value of $2790~\text{\AA}^3\,\text{atom}^{-1}$, the largest value in the dataset. However, we also collected two instances of the same polymer--coating combination oriented in the high-fluence Ram direction with $E_y$ values two or more orders of magnitude less than the Wake-facing measurement.

\subsection{Prompt Engineering and LLM fine-tuning}

\begin{figure}[H]
    \centering
    \includegraphics[width=1\linewidth]{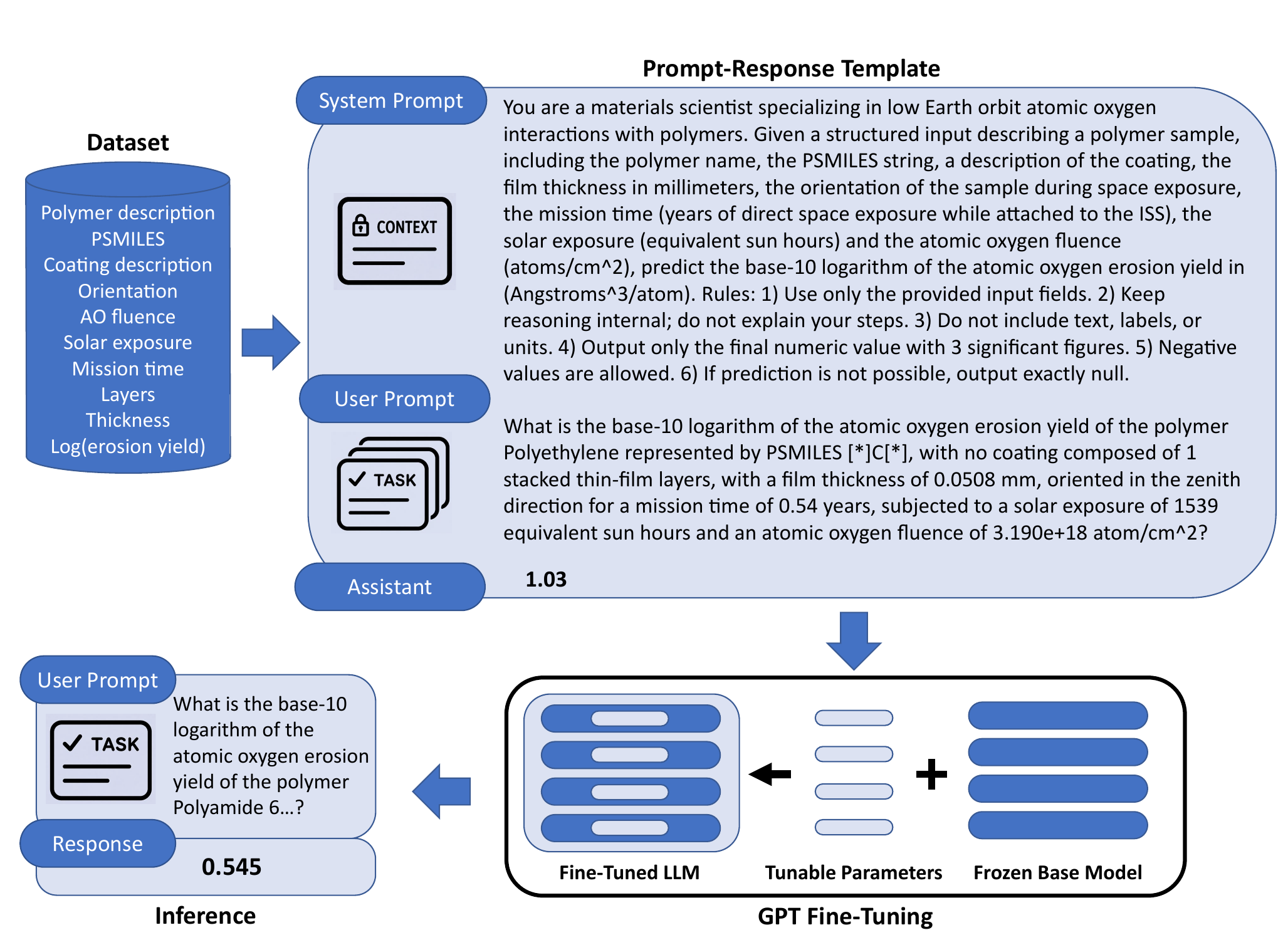}
    \caption{Fine-tuning workflow for erosion yield ($E_y$) prediction. The curated dataset, containing PSMILES, experimental conditions and log-transformed $E_y$ values, is divided into different splits. Prompt–response pairs are constructed with structured system and user prompts, and then used to fine-tune a base LLM. The resulting fine-tuned model is employed for inference to predict the $E_y$ of unseen polymer chemistries.}
    \label{fig3}
\end{figure}

We fine-tuned and tested GPT-4o using the dataset of 201 thin-film polymer samples and their measured $E_y$. To this end, information pertaining to a sample's chemical structure, macroscopic physical properties and the LEO environment during testing was systematically inserted into a prompt-response template. The template encouraged the model to adopt the role of an expert in LEO polymer-AO interactions with the ability to predict the $E_y$ of any polymer. In addition, a series of rules were provided that constrained the LLM’s behavior. These guidelines ensured the model always output a single number representing its prediction for the log-scaled $E_y$ in cubic angstroms per oxygen atom. An overview of the workflow used to fine-tune GPT-4o for the $E_y$ prediction task is shown in Figure~\ref{fig3}.

Each training example used for fine-tuning comprised three parts: the ``system prompt'', the ``user prompt'' and the ``assistant''. The system prompt provided the LLM with the general context in which the task was framed (Figure~\ref{fig3}). Furthermore, it was identical in every training instance, providing stability and consistency during learning. Conversely, the user prompt was unique to each example and presented the $E_y$ prediction task along with all relevant features including polymer description, PSMILES, coating description, orientation, AO fluence, solar exposure, mission duration, number of thin-film layers and the thickness of each layer as a natural language query. Finally, the assistant field provided the ground truth value, or simply the log-scaled experimental $E_y$ in cubic angstroms per incident AO atom. Since $E_y$ measurements span approximately seven orders of magnitude, a logarithmic transformation was applied to the $E_y$ targets to better capture relative variations and improve model stability.  During fine-tuning, the LLM's tunable parameters were updated to adapt it to the AO erosion task. After fine-tuning, the model was evaluated by providing it with novel prediction prompts. These were identical to training examples with the exception of a missing ground truth value (response), now replaced by the model's $E_y$ prediction. Finally, due to the stochastic nature of LLM outputs, each task was repeated 10 times, with the mean treated as the final prediction and the standard deviation as the uncertainty.

We evaluated model performance using two train--test splitting strategies: a random split and a more stringent restricted-group split. In the random split, the dataset was partitioned into training and test sets by simple random selection. As mentioned in the previous section, multiple measurements are represented by the same canonical PSMILES string, differing only in macroscopic configuration and environmental conditions. As a result, randomly generated splits often place the same polymer chemistry in both the training and test sets. Since resistance to AO erosion is fundamentally governed by polymer chemistry, the PSMILES string contains much of the predictive information and, in the absence of chemistry-based controls, constitutes a significant source of information leakage.

To address the shortcomings of the random split and better evaluate the model's ability to generalize to previously unseen polymer chemistries, we introduced the restricted-group split. In this approach, any polymer chemistry represented by multiple measurements was automatically assigned to the training set, ensuring that all test-set samples corresponded to chemistries absent during training. This resulted in chemically disjoint train--test pairs and a more rigorous assessment of the model's external chemical validity. Of the 201 measurements, only 29 satisfied this criterion and were therefore eligible for the restricted-group test sets. Notably, because all non-Ram measurements in the dataset correspond to polymers that were also tested in the Ram orientation, the restricted-group test sets consist exclusively of Ram-oriented samples. In this way, the restricted-group split primarily serves to assess the model's ability to generalize to new chemistries, while the random split gauges its ability to predict the behavior of familiar chemistries under different experimental conditions.

\subsection{Fine-tuning Optimization}

To obtain optimal performance, we refined GPT-4o’s hyperparameters, selecting for the most accurate $E_y$ predictions. Fine-tuning OpenAI's GPT models is particularly user-friendly; however, this accessibility comes at the cost of control. As a result, the only model variables tuned in this study were inference temperature and the number of training epochs. Inference temperature controls the model's creativity, with lower values producing more deterministic responses and higher ones boosting variability. The number of training epochs determines how many times the model is shown the entire training set during fine-tuning. To optimize both parameters, a series of fine-tuning experiments were conducted in which the effect of each variable was independently evaluated. Using five restricted-group train–test splits as an efficient yet exacting benchmark, the model was tasked with predicting the $E_y$ for 29 held-out polymer chemistries. These five restricted-group splits were used for all optimization steps because they provide the most rigorous assessment of model generalization while remaining computationally efficient and cost-effective. Lastly, every model's performance was quantified using the order of magnitude error (OME) and log-scaled coefficient of determination (log-R\textsuperscript{2}), defined in Equations~\ref{eq:ome} and~\ref{eq:logr2}

\begin{equation}\label{eq:ome}
    \text{OME} = \frac{1}{N} \sum_{i=1}^{N} \left| \log_{10}(y_i^{\text{true}}) - \log_{10}(y_i^{\text{pred}}) \right|
\end{equation}

\begin{equation}\label{eq:logr2}
    \log R^2 = 1 - \frac{\displaystyle\sum_{i=1}^{N} \left( \log_{10}(y_i^{\text{true}}) - \log_{10}(y_i^{\text{pred}}) \right)^2}{\displaystyle\sum_{i=1}^{N} \left( \log_{10}(y_i^{\text{true}}) - \overline{\log_{10}(y^{\text{true}})} \right)^2}
\end{equation}

\begin{figure}[H]
    \centering
    \includegraphics[width=1\linewidth]{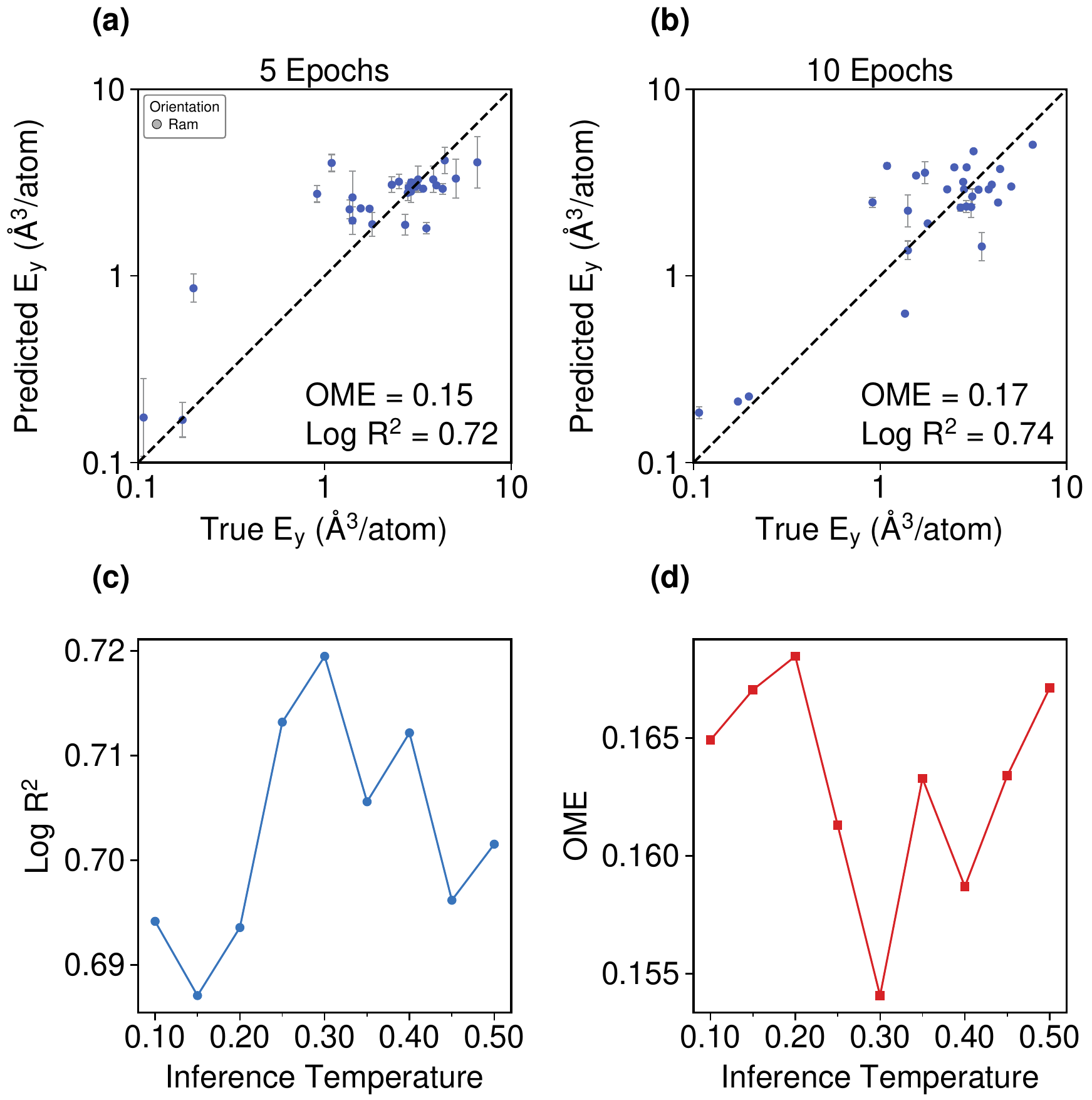}
    \caption{
    Effect of training epochs and inference temperature on GPT-4o performance. (a,b) Parity plots comparing predicted and experimental erosion yield ($E_y$) values after fine-tuning for 5 and 10 epochs, respectively, across five held-out test sets. The dashed line denotes perfect agreement ($y=x$). Each panel reports the corresponding order-of-magnitude error (OME) and log-R\textsuperscript{2}. (c,d) Dependence of log-R\textsuperscript{2} and OME, respectively, on the temperature used during inference.}
    \label{fig4}
\end{figure}

Figures~\ref{fig4}(a) and \ref{fig4}(b) present parity plots for GPT-4o after 5 and 10 fine-tuning epochs, respectively. Both visual inspection of these plots and a comparison of their corresponding performance metrics suggest training for 10 epochs produces no additional benefit. Accordingly, all remaining testing and validation were performed with models fine-tuned for 5 epochs. 
The values 5 and 10 were chosen based on a previous study in which GPT-3.5 was successfully trained to predict polymer solubility.\cite{doi:10.1021/acsmaterialslett.5c00054} The same paper further concluded that an inference temperature of 0.5 was ideal for such a task. In the present study, however, model performance peaked at an inference temperature of 0.3, as shown in Figure~\ref{fig4}(c) and \ref{fig4}(d), supporting its selection for future $E_y$ predictions.

\begin{figure}[H]
    \centering
    \includegraphics[width=1\linewidth]{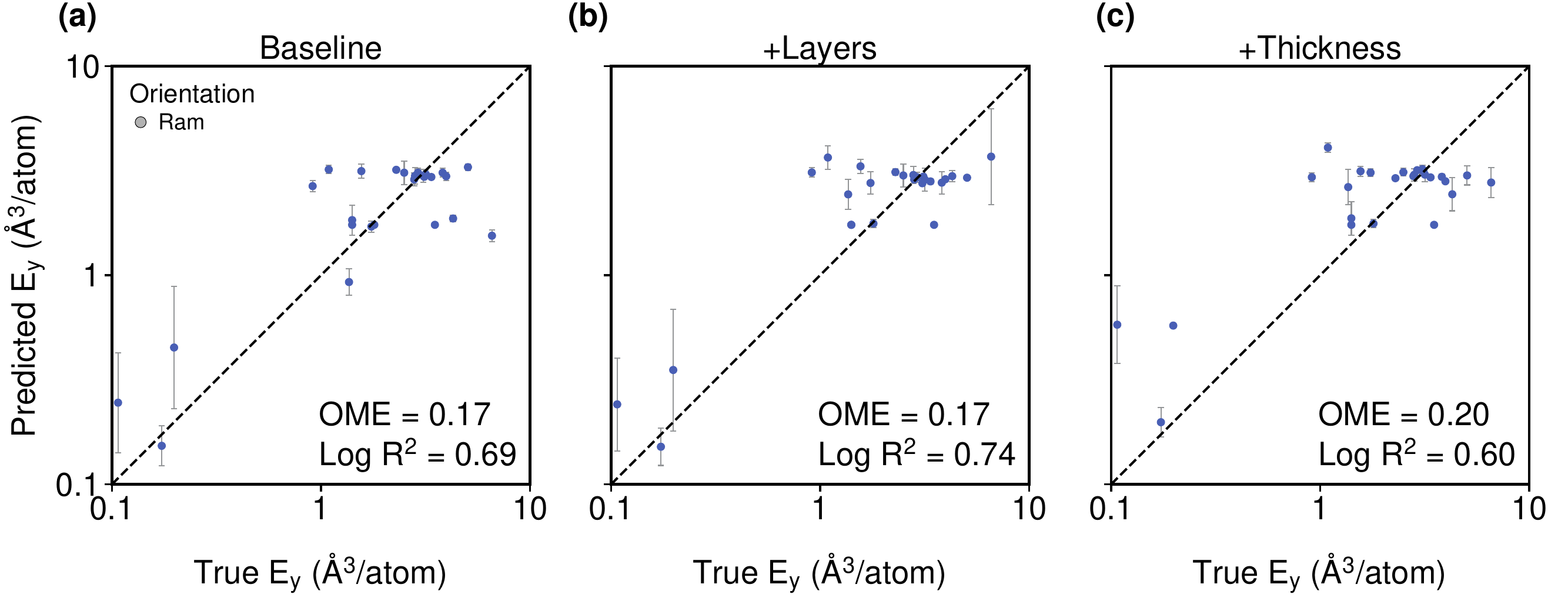}
    \caption{Effect of macroscopic features on model performance. Parity plots comparing predicted and true erosion yield ($E_y$) values for models trained (a) without thickness and number of layers, (b) with number of layers and (c) with thickness as a feature, across five train-test splits.}
    \label{fig5}
\end{figure}

Following inference temperature and epoch tuning, we assessed the importance of macroscopic structural descriptors in predicting $E_y$, namely the number of thin-film layers and the thickness of each layer. Because 32 entries lack values for both features, the ablation was restricted to samples where both were known. We fine-tuned OpenAI GPT-4o using three distinct prompt–response templates and evaluated performance on 26 unseen chemistries across the same five restricted-group test sets used in the preceding evaluation, excluding three samples with missing layer and thickness data. The first template (Figure~\ref{fig5}(a)) acted as a control, omitting both ``layers'' and ``thickness''. This was followed by a version including ``layers'' (Figure~\ref{fig5}(b)) and another including ``thickness'' (Figure~\ref{fig5}(c)). Both the OME and log-R\textsuperscript{2} values remained largely consistent following the inclusion of these two macroscopic descriptors. This suggests that a polymer’s suitability for the LEO environment is primarily governed by its chemistry rather than macroscopic film characteristics, justifying the omission of both number of layers and per-layer thickness from all subsequent training and evaluation.

\begin{figure}[H]
    \centering
    \includegraphics[width=1\linewidth]{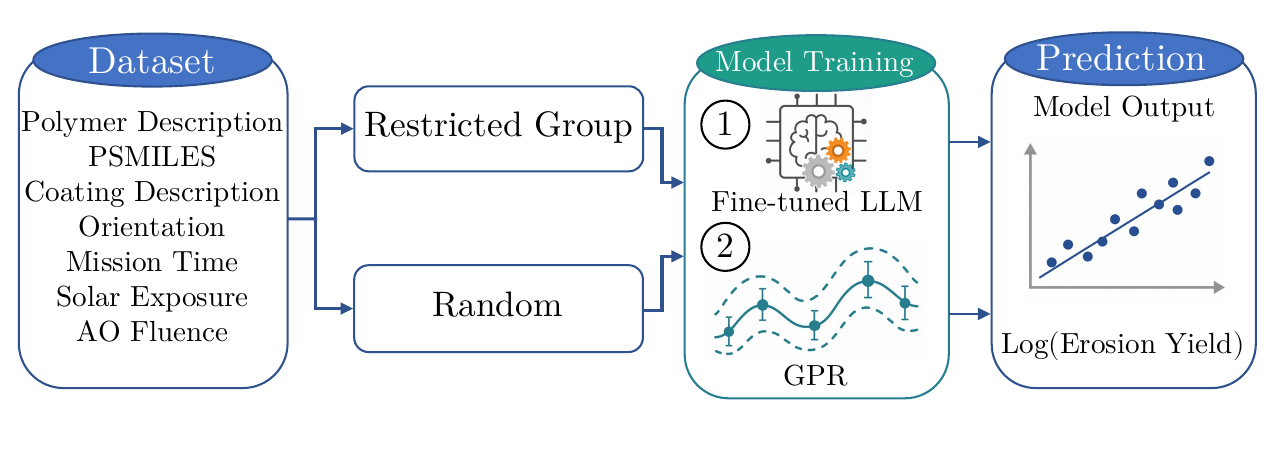}
    \caption{Workflow for evaluating the fine-tuned GPT-4o model and benchmarking it against Gaussian process regression (GPR). The dataset is partitioned using random and restricted-group train--test splits to assess interpolation across known chemistries and generalization to unseen polymer chemistries, respectively. Both models were trained on the same experimental measurements but differed in how the polymer and associated descriptors were represented and encoded.}
    \label{fig6}
\end{figure}

The preceding experiments employed five restricted-group train--test splits for simplicity and computational efficiency. However, a comprehensive evaluation of the fully optimized LLM requires the use of both the random and restricted-group splitting methods (Figure~\ref{fig6}). Together, the two strategies assess complementary aspects of model performance: interpolation across previously seen chemistries under different environmental conditions, and extrapolation to entirely unseen polymer chemistries, respectively. Furthermore, to benchmark our approach against a conventional machine learning method, we trained a Gaussian process regression (GPR) model (Figure~\ref{fig6}) using loop-closed Morgan fingerprints\cite{rogers2010ecfp} and a radial basis function (RBF) kernel with the fingerprint radius, length, the RBF length scale, kernel amplitudes, and noise level optimized during model fitting. The GPR model was evaluated on the same random and restricted-group splits as the fine-tuned GPT-4o model. Table~\ref{tab:features} compares the input descriptors used by the two approaches. For GPT-4o, the polymer repeat unit was represented by its raw PSMILES string, while the GPR model employed Morgan fingerprints. Likewise, the continuous descriptors (AO fluence, solar exposure, and mission time) were provided directly to GPT-4o but were log-transformed and standardized for GPR, while the categorical descriptors, such as orientation and coating description, were provided as natural language to GPT-4o but required one-hot encoding for GPR. Lastly, the polymer description was included only for GPT-4o, as the unstructured textual data it contains cannot be readily captured by the numerical representation used in GPR. These differences highlight one of the principal advantages of LLMs: their ability to directly incorporate heterogeneous textual data without extensive feature engineering.

\begin{table}[H]
\centering
\small
\caption{Comparison of the input representations used by the fine-tuned GPT-4o model and the Gaussian process regression (GPR) benchmark. Both models were trained on the same experimental measurements but differed in how the polymer and associated descriptors were represented and encoded.}
\label{tab:features}
\begin{tabular}{@{}lll@{}}
\toprule
\textbf{Feature} & \textbf{LLM} & \textbf{GPR} \\
\midrule
PSMILES (repeat unit) & \checkmark\ (PSMILES string)          & \checkmark\ (Morgan fingerprint; Tanimoto kernel) \\
AO fluence            & \checkmark\ (atoms\,cm$^{-2}$)       & \checkmark\ ($\log_{10}$, standardized; RBF kernel) \\
Solar exposure        & \checkmark\ (ESH)                    & \checkmark\ ($\log_{10}$, standardized; RBF kernel) \\
Mission time          & \checkmark\ (yr)                     & \checkmark\ ($\log_{10}$, standardized; RBF kernel) \\
Orientation           & \checkmark\ (Ram/Wake/Zenith/Nadir)  & \checkmark\ (one-hot; RBF kernel) \\
Coating description    & \checkmark\ (textual)            & \checkmark\ (one-hot; RBF kernel) \\
Polymer description   & \checkmark\ (textual)                & $\times$ \\
\bottomrule
\end{tabular}
\end{table}

\begin{figure}[H]
    \centering
    \includegraphics[width=1\linewidth]{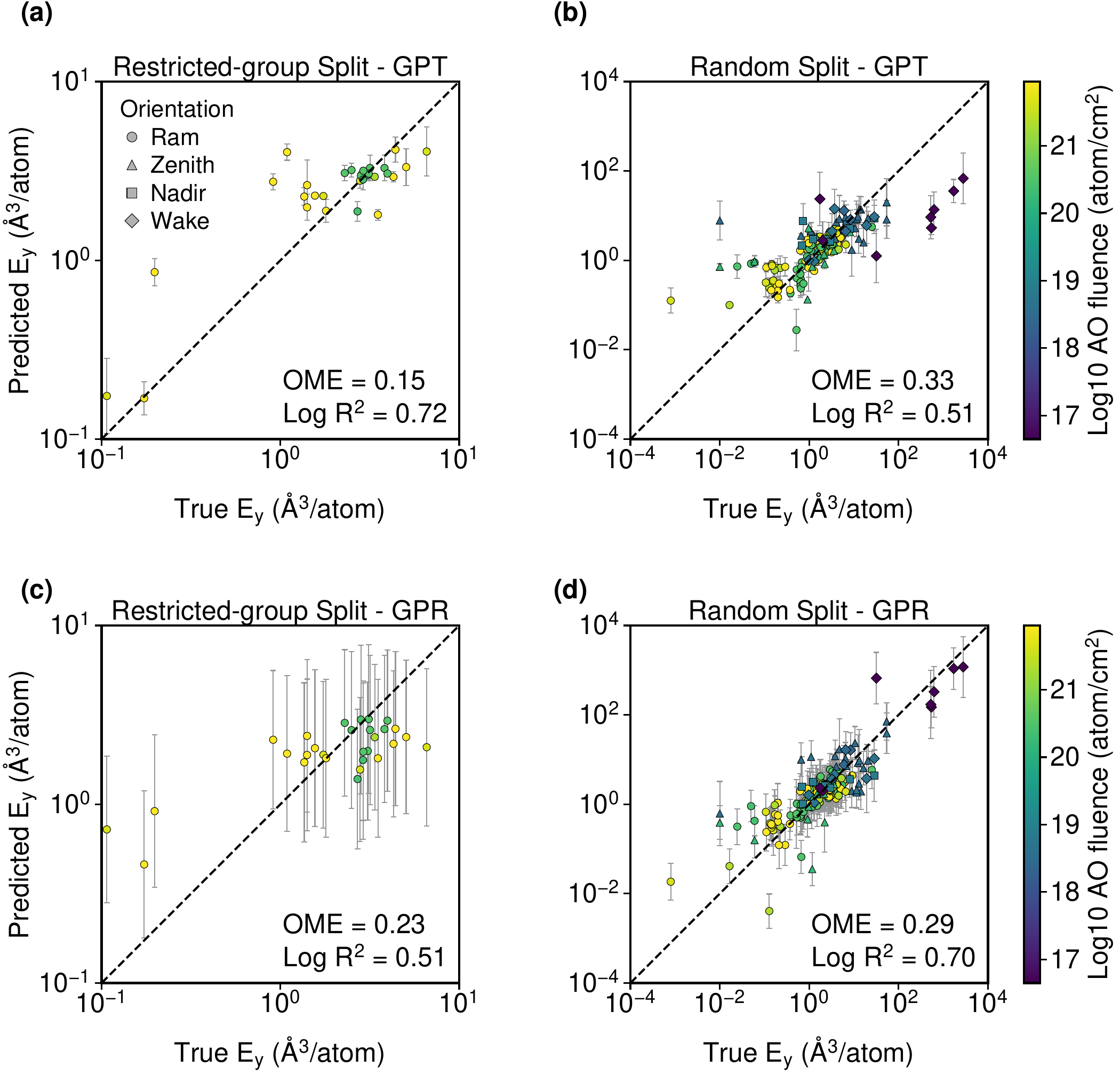}
    \caption{A comparison of model performance across data splitting strategies and informatics methods. Parity plots of predicted versus true erosion yield ($E_y$) are shown for fine-tuned GPT-4o using (a) the restricted-group split, containing only Ram-oriented measurements of unseen chemistries, and (b) the random split, including all four orientations, and for GPR using (c) the restricted-group split and (d) the random split. Data points are colored by AO fluence, illustrating model performance across different exposure conditions and highlighting the relationship between AO fluence, sample orientation, and erosion yield. The final GPT-4o models were fine-tuned with optimized hyperparameters.}
    \label{fig7}
\end{figure}

The parity plots for fine-tuned GPT-4o and the GPR model under both splitting strategies are shown in Figure~\ref{fig7}(a-d). Data points are color coded by fluence to show model performance under different experimental conditions while also demonstrating the relationship between fluence, orientation and erosion yield as discussed in the data analysis section. Comparison of their corresponding OME and log-R\textsuperscript{2} values shows that the fine-tuned LLM clearly outperformed GPR on the restricted-group split, achieving an OME of 0.154 and a log-R\textsuperscript{2} of 0.719. GPT-4o's superior performance under the restricted-group split indicates a stronger ability to generalize to previously unseen polymer chemistries and therefore a greater aptitude for screening novel materials. Nevertheless, given the training data's limited chemical diversity, predictions involving entirely new or underrepresented chemistries should be interpreted with caution. Meanwhile, under the random split, the fine-tuned LLM was outperformed by GPR. However, stratifying the random split predictions by fluence reveals that the lowest fluence bin ($<10^{18}$ atoms\,cm$^{-2}$) is entirely responsible for this difference. This suggests GPR was better able to capture $E_y$ trends under less severe LEO conditions. However, since a key criterion for next-generation space polymers is the ability to withstand high fluences, GPR's superior performance in the low-fluence regime is of limited utility.

One noteworthy aspect of Figure~\ref{fig7} is the greater accuracy observed under the restricted-group split despite it being the more stringent test of the LLM's abilities. This discrepancy is explained by outlier $E_y$ values in the random split test sets that were ineligible for the restricted-group test sets. Concretely, the restricted-group split only tests GPT-4o on polymers with $E_y$ values between 0.1 and 10 cubic angstroms per incident atom, a range containing most of the training data. The random split, by contrast, includes polymers with true $E_y$ values spanning roughly 0.0001 to 1000. Consequently, the random split asks the model to make predictions in regions where there is little to no training data, substantially degrading performance. 

The fine-tuned LLM's overall superior performance is partly attributed to its innate ability to handle and extract meaning from the natural language contained in the polymer descriptions. GPR models are fundamentally ill-equipped to handle textual data, and their ability to incorporate categorical features is of little use when most of the polymer descriptions are entirely unique. Another likely reason for GPT-4o’s superior performance is its advantage in the low-data regime. When data is sparse, traditional models, such as GPR, often struggle to extract reliable trends from limited examples. LLMs, by contrast, are better suited to extrapolate from small datasets by combining patterns in the training data with knowledge acquired during pretraining. We theorize that this prior knowledge may help the model generalize more effectively to novel polymer chemistries.

%\begin{figure}[H]
 %   \centering
%    \includegraphics[width=1\linewidth]{images/gpt4o_vs_gpr_zoomed.eps}
 %   \caption{Parity plots comparing the performance of fine-tuned GPT-4o and GPR on the erosion yield prediction task.}
 %   \label{fig1}
%\end{figure}

\section{Outlook}

Given the extreme scarcity of $E_y$ measurements under genuine LEO conditions, the fine-tuned LLM is the current architecture of choice for space polymer informatics. As additional LEO data become available, the opportunity to explore a broad range of machine learning approaches and hybrid modeling strategies can offer a means of both improving predictive performance and generating further insight into the AO erosion phenomenon. In particular, data-fusion techniques, such as multitask learning, offer a promising route for integrating scarce LEO measurements with complementary ground-based experiments and simulation-derived datasets. An integrated framework of this sort would allow models to benefit from both high-fidelity space exposure data and more abundant surrogate data sources.\cite{KUENNETH2021100238,toland2023accelerated,phan2024gas}

To this end, future LEO experiments should prioritize chemical diversity over variations on previously characterized polymers, enabling more informed chemical informatics. To complement this, ground-based testing and simulation work should focus on generating data that can be correlated with existing high-fidelity LEO experiments. A sufficient collection of such data would help establish a clear quantitative bridge between experiments conducted in the relatively inaccessible LEO environment and more readily available terrestrial approaches.

\section{Conclusions}

In this study, we developed a data-driven framework for predicting the atomic oxygen erosion yield ($E_y$) of polymers in low Earth orbit (LEO) environments using a fine-tuned large language model (LLM). By leveraging experimental data from multiple NASA missions and incorporating relevant exposure conditions, the LLM demonstrates reasonable predictive capabilities, even in the low-data regime. During the fine-tuning process, we found that GPT-4o's predictions were unaided by the presence of macroscopic descriptors such as film thickness and layering. This suggests that the AO erosion phenomenon is largely governed by environmental and chemical considerations rather than materials architecture, highlighting the importance of molecular-level design in developing AO-resistant materials. Additionally, GPT-4o consistently outperformed a traditional machine learning approach under situations relevant to space materials design without complex feature engineering. This approach offers a scalable and efficient framework for screening polymer candidates for space applications, reducing the reliance on costly and time-intensive experimental testing.

\section{Data and Code Availability}

The data and source code used in this study are available in the GitHub repository: \href{https://github.com/Ramprasad-Group/polymer-erosion-yield}{polymer-erosion-yield}.

\begin{suppinfo}
Dictionary of the 55 polymer names used in Figure~\ref{fig2}(a) and their corresponding canonical PSMILES representations.
\end{suppinfo}

\begin{acknowledgement}
The authors gratefully acknowledge the SPACEMAT University Center of Excellence (UCoE), funded by Air Force Research Laboratory (AFRL) under contract number AFRL FA8650-24-2-5210, for supporting this work. The authors also thank Kim de Groh of NASA Glenn Research Center as well as Elena Plis and Jud Ready of Georgia Tech Research Institute (GTRI) for helpful discussions.
\end{acknowledgement}

% %%%%%%%%%%%%%%%%%%%%%%%%%%%%%%%%%%%%%%%%%%%%%%%%%%%%%%%%%%%%%%%%%%%%%
% %% The same is true for Supporting Information, which should use the
% %% suppinfo environment.
% %%%%%%%%%%%%%%%%%%%%%%%%%%%%%%%%%%%%%%%%%%%%%%%%%%%%%%%%%%%%%%%%%%%%%
% \begin{suppinfo}

% This will usually read something like: ``Experimental procedures and
% characterization data for all new compounds. The class will
% automatically add a sentence pointing to the information on-line:

% \end{suppinfo}

\bibliography{acs-achemso.bib}

\end{document}